# Phase control in La-214 epitaxial thin films

Michio Naito, Akio Tsukada[a], Tine Greibe, and Hisashi Sato[b]

NTT Basic Research Laboratories, NTT Corporation,
3-1, Morinosato-Wakamiya, Atsugi-shi, Kanagawa 243-0198, Japan

[a]Department of Physics, Science University of Tokyo,
1-3, Kagurazaka, Shinjuku-ku, Tokyo 162-8601, Japan

[b]NTT Network Innovation Laboratories, NTT Corporation,
1-1, Hikarino-oka, Yokosuka-shi, Kanagawa 239-0847, Japan

## ABSTRACT

The lanthanide (Ln) copper oxides of the general chemical formula $Ln_2CuO_4$ take two different crystal structures: $K_2NiF_4$ (T) and $Nd_2CuO_4$ (T'). $La_2CuO_4$ takes the T structure by high-temperature bulk processes. The "thermal expansion mismatch" between the La-O and Cu-O bonds predicts that the T' phase of $La_2CuO_4$ can be stabilized at synthesis temperatures below 425°C. Such low synthesis temperatures are difficult to access by bulk processes, but easy by thin-film processes. We have surveyed growth parameters in molecular beam epitaxy, and succeeded in the selective stabilization of T- and T'-$La_2CuO_4$. From our observations, it turns out that the growth temperature as well as the substrate play a crucial role in the selective stabilization: the T' structure is stabilized at low growth temperatures (< 600°C) and with substrates of $a_s$ < 3.70 Å or $a_s$ > 3.90 Å, while the T structure is stabilized at high growth temperatures (> 650°C) or with substrates of $a_s$ ~ 3.70 - 3.85 Å. We have also been attempting hole (Ca, Sr, and Ba) and electron (Ce) doping into both of T- and T'-$La_2CuO_4$. In T-$La_2CuO_4$, hole doping produces the well-known LSCO and LBCO. Surprisingly, contrary to the empirical law, electron doping is also possible up to $x$ ~ 0.06 - 0.08, although the films do not show superconductivity. In T'-$La_2CuO_4$, electron doping produces superconducting T'-$(La,Ce)_2CuO_4$ with $T_c$ ~ 30 K, although hole doping has as yet been unsuccessful.

---

Further author information:
M. N. (correspondence): E-mail: michio@will.brl.ntt.co.jp; Web: http://www.brl.ntt.co.jp/people/michio/index.html
Telephone: +81-46-240-3516; FAX: +81-46-240-4717
A. T. : E-mail: tsukada@with.brl.ntt.co.jp; Telephone: +81-46-240-3349
T. G. : On leave from Technical University of Denmark.
H. S. : E-mail: satoh@wslab.ntt.co.jp; Telephone: +81-468-59-3467; FAX: +81-468-55-1497

# 1. INTRODUCTION

The lanthanide copper oxides of the general chemical formula $La_2CuO_4$ possess a richness of structural and physical properties because of the wide range of lanthanide (Ln) ion sizes. There are two closely related structures as shown in Fig. 1: $K_2NiF_4$ (T) and $Nd_2CuO_4$ (T'). The structural difference between T and T' can be viewed simply as the difference in the Ln-O arrangements: rock-salt-like versus fluorite-like. The key parameter determining which of these structures is formed is the ionic radius of the $Ln^{3+}$ ion. The T structure is formed with large $La^{3+}$, while the T' structure is formed with smaller $Ln^{3+}$ ions, such as Ln = Pr, Nd, Sm, Eu, and Gd.[1-3] In high-pressure synthesis, the T' structure can also be formed by Dy to Tm as well as Y.[4] The T-T' boundary lies between $La^{3+}$ and $Pr^{3+}$. In the mixed-lanthanide system $(La,Ln)_2CuO_4$, a third structure (denoted T*), which consists of alternate stacks of T and T' slabs, is observed.[5] The stabilization of this structure requires two Ln ions of significantly different size: one large $La^{3+}$ and a much smaller $Ln^{3+}$.

$La_2CuO_4$ takes the T structure by high-temperature bulk processes. However, $La_2CuO_4$ is at the borderline stability of the T structure. It has been predicted that the T' phase of $La_2CuO_4$ can be stabilized at synthesis temperatures below 425°C because of the thermal expansion difference between the La-O and Cu-O bonds.[6] Such low synthesis temperatures are difficult to access by bulk processes, but easy by thin-film processes. In this article we survey growth parameters in molecular beam epitaxy toward the selective stabilization of T- and T'-$La_2CuO_4$.

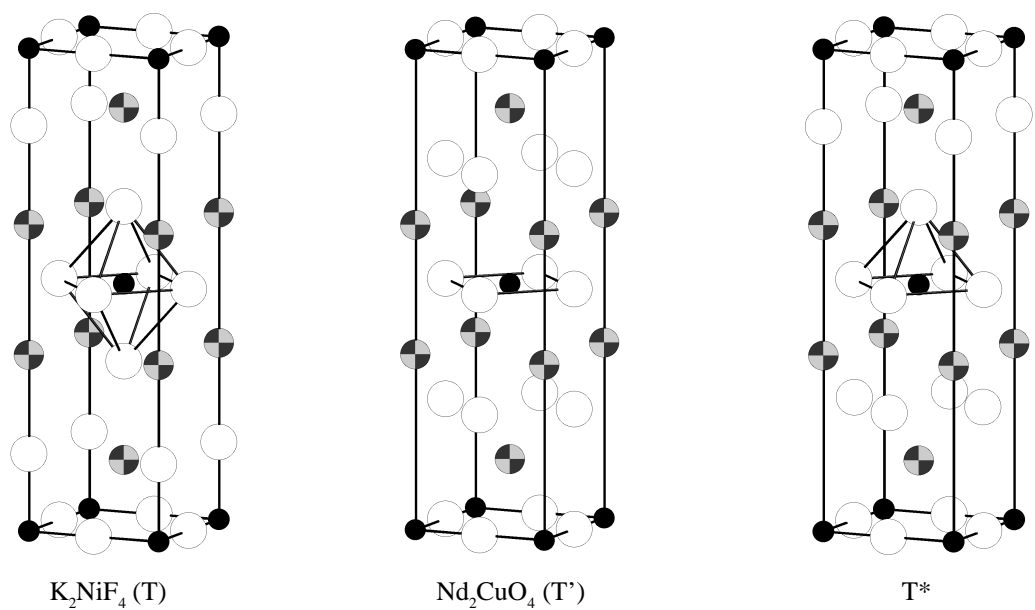

$K_2NiF_4$ (T)  $Nd_2CuO_4$ (T')  T*

Figure 1. Three crystal structures of lanthanide cuprates $(Ln_2CuO_4)$. The $K_2NiF_4$ (T) structure has octahedral $CuO_6$, the $Nd_2CuO_4$ (T') structure has square-planar $CuO_4$, and the T* structure has pyramidal $CuO_5$.

## 2. REVIEW ON THE PAST WORKS ON 214 CUPRATES

### 2.1 Tolerance factor

The crystal chemistry in the lanthanide copper oxides has been explained by Bringley et al.[2] and Manthiram et al.[3] in terms of the crystallographic (Goldshimidt) tolerance factor ($t$), which is defined as

$$t = \frac{r_i(\text{Ln}^{3+}) + r_i(\text{O}^{2-})}{\sqrt{2} \times (r_i(\text{Cu}^{2+}) + r_i(\text{O}^{2-}))} \;, \tag{1}$$

where $r_i(\text{Ln}^{3+})$, $r_i(\text{Cu}^{2+})$, and $r_i(\text{O}^{2-})$ are the empirical room-temperature ionic radii for $\text{Ln}^{3+}$, $\text{Cu}^{2+}$, and $\text{O}^{2-}$ ions by Shannon and Prewitt.[7] The tolerance factor ($t$) was initially proposed to argue the stability of the perovskite structure ($\text{ABO}_3$).[8] It represents the bond length matching between AO layers and $\text{BO}_2$ layers. Ideal matching corresponds to $t = 1$, and the perovskite structure is stable within ~ $0.8 < t < 1.0$. This factor has also been used to explain the stability fields of the T, T', and T* structures in the $\text{A}_2\text{BO}_4$ stoichiometry. Figure 2 is a schematic illustration of the bond length mismatch between Cu-O and La-O in the case of $\text{La}_2\text{CuO}_4$. In calculating the tolerance factor in eq. (1), one should use $r_i(\text{Ln}^{3+})$ for nine-fold Ln-O coordination corresponding to the T structure.[2] This is because the Ln-O coordination of the T structure is the most suitable reference state, for which $t = 1$ represents ideal matching as in the case of the perovskite structure. The calculated values for $t$ are listed in Table 1.

On the basis of a systematic investigation of the mixed lanthanide system $\text{La}_{2-x}\text{Ln}_x\text{CuO}_4$, in which "average" $t$ can be varied continuously, Bringley et al.[2] found that the T structure is stable for $0.87 \leq t \leq 0.99$, the T' structure is stable for $0.83 \leq t \leq 0.86$, and neither T nor T' is formed for $t < 0.83$. When $t < 0.83$, the complicated so-called "$\text{Ho}_2\text{Cu}_2\text{O}_5$" structure with six-fold Ln-O coordination is formed instead.[9] These results can be interpreted as follows. When $t$ is close to 1, i.e., when the bond length matching is near ideal, the T structure is stable. For $t$ substantially smaller than 1, the T phase becomes unstable. The first indication of T-phase instability is the occur-rence of orthorhombic distortion as seen in $La_2CuO_4$ ($t \sim 0.868$). The distortion occurs so as to accommodate the large bond length mismatch by tilting of $CuO_6$ octahedra. For $t < 0.86$, the bond length mismatch becomes untolerable, resulting in transformation to the T' phase. The critical value for the T  T' transition is $t_c = 0.865$. As regards the T* phase, its occurrence is limited to a very narrow range of $t$: $0.85  t  0.86$. Furthermore, the stabilization of pure T* requires sufficient disparity in the ionic size between $La^{3+}$ and its counter lanthanide cation $1.12  r_i(La^{3+})/r_i(Ln^{3+})  1.19$.[2]

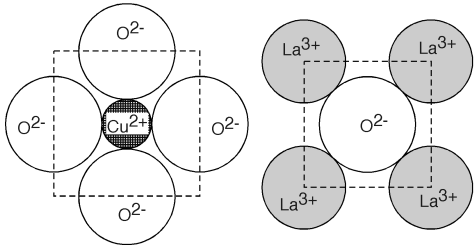

Figure 2. Sliced views of $\text{CuO}_2$ and LaO layers in $\text{La}_2\text{CuO}_4$. The cell size of $\text{CuO}_2$ layer is substantially larger than that of LaO layers. The tolerance factor $t$ is close to the critical $t_c$ for the T $\leftrightarrow$ T' transition.

Table 1. Ionic radius for nine-fold coordination $\text{Ln}^{3+}$ and the tolerance factor calculated from eq. (1) with $r_i(\text{Cu}^{2+}(\text{VI})) = 0.73$ Å and $r_i(\text{O}^{2-}(\text{VI})) = 1.40$ Å.

| $\text{Ln}^{3+}$ | $r_i(\text{Ln}^{3+}(\text{IX}))$ [Å] | tolerance factor $t$ |
|---|---|---|
| $\text{La}^{3+}$ | 1.216 | 0.8685 |
| $\text{Ce}^{3+}$ | 1.196 | 0.8618 |
| $\text{Pr}^{3+}$ | 1.179 | 0.8562 |
| $\text{Nd}^{3+}$ | 1.163 | 0.8509 |
| $\text{Pm}^{3+}$ | 1.144 | 0.8445 |
| $\text{Sm}^{3+}$ | 1.132 | 0.8406 |
| $\text{Eu}^{3+}$ | 1.120 | 0.8366 |
| $\text{Gd}^{3+}$ | 1.107 | 0.8323 |
| $\text{Tb}^{3+}$ | 1.095 | 0.8283 |
| $\text{Dy}^{3+}$ | 1.083 | 0.8243 |
| $\text{Ho}^{3+}$ | 1.072 | 0.8206 |
| $\text{Er}^{3+}$ | 1.062 | 0.8173 |
| $\text{Tm}^{3+}$ | 1.052 | 0.8140 |
| $\text{Yb}^{3+}$ | 1.042 | 0.8107 |
| $\text{Lu}^{3+}$ | 1.032 | 0.8074 |
| $\text{Y}^{3+}$ | 1.075 | 0.8216 |

**2.2 Thermal-expansion mismatch**

The above discussion neglects the temperature effect on the bond length. As pointed out by Manthiram and Goodenough,[6] the different thermal expansion ("thermal-expansion mismatch") of the Ln-O and Cu-O bond length plays an important role in the T-versus-T' stability. They claimed that the "ionic" Ln-O bond has larger thermal expansion than the "covalent" Cu-O bond, and that this leads to the increase of $t$ with increasing temperature. One manifestation of this temperature effect is the structural phase transition observed in $La_2CuO_4$. For $La_2CuO_4$, $t$ is 0.8685 at room temperature, but $t$ may get close to 0.9 at 1000°C. This causes the low-temperature orthorhombic (LTO) to high-temperature tetragonal (HTT) structural transition at around 550 K. For $La_{2-x}Sr_xCuO_4$ (LSCO), the room-temperature tolerance factor increases with the Sr doping level ($x$), since $r_i(Sr^{2+})$ is larger than $r_i(La^{3+})$ and more importantly the Cu-O bond shrinks by hole doping. Then, as schematically illustrated in Fig. 3, the HTT structure is maintained down to lower temperature for higher $x$. Namely, the LTO-to-HTT phase transition temperature decreases with $x$.[10, 11]

Another interesting indication of the thermal expansion mismatch is the possibility of the phase control between T and T' by changing the synthesis temperature. Manthiram and Goodenough demonstrated the success in the selective stabilization of T versus T' in the $La_{2-y}Nd_yCuO_4$ system by changing the firing temperature ($T_s$) from 500°C to 1050°C.[6] $La_{1.5}Nd_{0.5}CuO_4$ can be stabilized as single-phase T' below $T_s = 625°C$ or single-phase T at $T_s = 775 - 850°C$. A two-phase mixture of T and T' is obtained between 625°C and 775°C, and a new phase (T'' by their notation) appears above 950°C. In their experiments, coprecipitated fine powders of hydroxides/carbonates were used to promote chemical reaction at firing temperatures as low as 500°C. At $T_s$ = 500°C, even $La_2CuO_4$ becomes not single-phase T but a two-phase mixture of T and T'. By extrapolation of the T/T' phase boundary in the $La_{2-y}Nd_yCuO_4$ system to $y = 0$, they predicted $La_2CuO_4$ can be stabilized in the T' structure below $T_s$ = 425°C. Since a firing temperature of at least 500°C was required in their bulk process even when coprecipitated powders were used, they could not achieve single-phase T'-$La_2CuO_4$. The synthesis of T'-$La_2CuO_4$ by bulk synthesis has been achieved only by a very special recipe as given by Chou et al.[12] The recipe consists of the following two steps. The first step is to reduce T-$La_2CuO_4$ with hydrogen around 300°C and to obtain the $Sr_2CuO_3$-like phase. The second step is to convert the $Sr_2CuO_3$-like phase to T'-$La_2CuO_4$ by reoxygenation below 400°C. The resultant product shows single-phase T', although X-ray peaks are broadened due to considerable lattice disorder and defects.

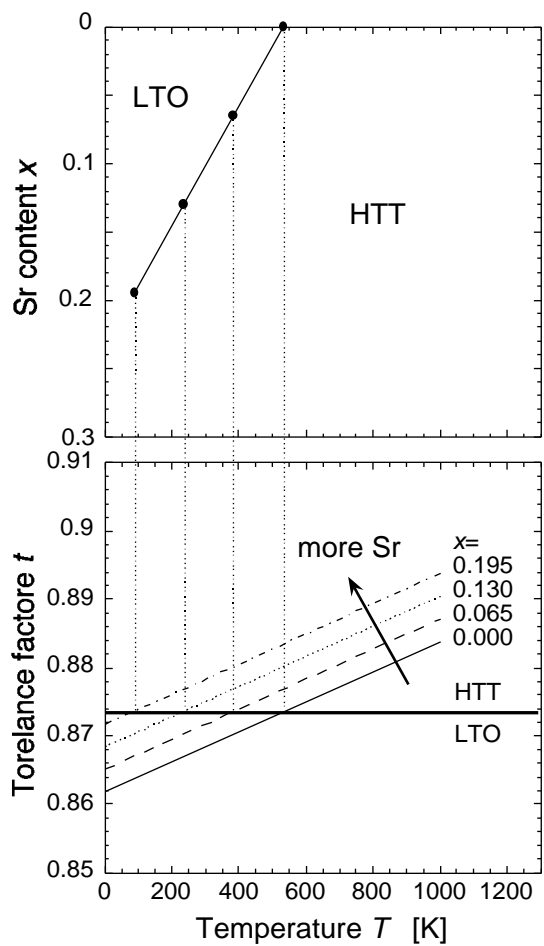

Figure 3. *Lower*: temperature dependences (schematic) of the tolerance factor ($t$) in $La_{2-x}Sr_xCuO_4$ systems with $x = 0.00, 0.065, 0.13$, and 0.195. The horizontal line represent the HTT-LTO phase boundary. The substitution of $Sr^{2+}$ for $La^{3+}$ increases the tolerance factor, and thereby stabilize the HTT structure to lower temperature. *Upper*: resultant $x$ dependence of the HTT-LTO transition temperature in $La_{2-x}Sr_xCuO_4$.

By using thin-film synthesis methods, the reaction temperature can be lowered significantly, since reactants are much smaller in size and more reactive than in bulk synthesis. The reactants in thin film synthesis are atoms or molecules or ions or clusters, depending on the technique employed. The limiting case is achieved by reactive coevaporation from metal sources, in which the reactants are atoms and the oxidation reaction is initiated on a substrate. Using this reactive coevaporation technique, we have learned from our ten-year experience that cuprate films crystallize at temperatures as low as 400°C. This means that the survey of the phase diagram in 214 cuprates can be extended down to ~ 400°C, which enabled us to stabilize the T' phase of pure La-214.

**2.3 Doping and superconductivity in 214 cuprates**

The T and T' structures are similar, but they show significant differences in many respects. The T structure has $CuO_6$ octahedra and is empirically supposed to accept only hole doping, which leads to $p$-type superconductivity.[13] The most typical example of this is $(La,Sr)_2CuO_4$ (LSCO) with optimum $T_c$ of 37 K.[14] The T' structure has two-dimensional square-planar $CuO_4$ and is supposed to accept only electron doping, which leads to $n$-type superconductivity.[13] Typical examples are $(Pr,Ce)_2CuO_4$ (PCCO) and $(Nd,Ce)_2CuO_4$ (NCCO) with optimum $T_c$ of 25 K.[15, 16]

As mentioned above, the tolerance factor $t$ of $La_2CuO_4$ is close to the critical $t_c$ of the T  T' transition. Therefore, the substitution of $La^{3+}$ by smaller $Ce^{4+}$ destabilizes the T phase, and stabilizes the T' phase. In fact, this phase change with Ce doping was observed in 1989 by Takayama-Muromachi et al.[17] They obtained T'-$La_{1.85}Ce_{0.15}CuO_4$ at the firing temperature of around 600°C using coprecipitated powder. However, their specimens did not show superconductivity even after the reduction heat treatment that is a prerequisite to achieving $n$-type superconductivity in the T' cuprates. The superconductivity at $T_c$ ~ 30 K in T'-$(La,Ce)_2CuO_4$ was first confirmed in 1994 by Yamada et al.[18] Their specimens were also prepared at around 600°C by a special precursor technique. These previous bulk works seem to indicate that the upper limit of the firing temperature to obtain T'-$La_{2-x}Ce_xCuO_4$ ($x$ ~ 0.15) is around 600°C, which is substantially increased (by 175°C) compared with $T_s$ = 425°C to stabilize the T' phase for pure $La_2CuO_4$. The firing temperature of 600°C, however, is still very low for bulk synthesis. The resultant specimens are not well crystallized and the superconducting properties are poor. Furthermore, the single-phase can be obtained only in a very narrow range of $x$ ($0.13 < x < 0.15$). In contrast, the growth temperature of 600°C is sufficiently high to get well crystallized cuprate films. In fact, it was recently shown by Matsuo et al.[19] that pulsed laser deposition can easily produce T'-$La_{2-x}Ce_xCuO_4$ epitaxial films for a wide range of $x$ on $SrTiO_3$ substrates, although their specimens were not superconducting.

## 3. MBE GROWTH

We grew La-214 thin films in a customer-designed MBE chamber from metal sources using multiple electron-gun evaporators with accurate stoichiometry control of the atomic beam fluxes. During growth, RF activated atomic oxygen or ozone was used for oxidation. The growth either with RF activated atomic oxygen or with ozone produces essentially the same quality of films at $T_s$ above 600°C, but at $T_s$ below 600°C the growth with atomic oxygen seems to give better results. So most of the films grown below $T_s$ = 600°C were prepared with atomic oxygen. The growth rate was ~1.5 Å/sec, and the film thickness was typically 500 - 1000 Å. In order to examine the substrate influence on the selective phase stabilization, we used various substrates as listed in Table 2. The in-plane lattice constant ($a_s$) covers 3.6 Å to 4.2 Å, which should be compared with $a_0$ = 3.803 Å for T-$La_2CuO_4$ or $a_0$ = 4.005 Å for T'-$La_2CuO_4$. The crystal structures include perovskite, $K_2NiF_4$, NaCl, and $CaF_2$ (fluorite). We deposited films simultaneously on all the substrates listed in Table 2, which were pasted to one substrate holder by Ag paint. This avoids run-to-run variations and saves time.

Hole doping was attempted either by substituting divalent Ba, Sr, or Ca for trivalent La, or introducing extra oxygen.[20,21] Electron doping was attempted by substituting tetravalent Ce for La.[20] For electron-doped T' films, apical oxygen, which is harmful to superconductivity, was removed by heat treatment at around 600°C in vacuum. The composition was calibrated by inductively coupled plasma spectroscopy (ICP).

Table 2. Crystal structure and $a$-axis parameter ($a_s$) for the substrates used in this work. The in-plane lattice constants for T'-$La_2CuO_4$ and T-$La_2CuO_4$ are also included.

| substrate | abbreviation | $a_s$ [Å] | crystal structure |
|---|---|---|---|
| MgO (100) | MgO | 4.212 | NaCl |
| $KTaO_3$ (100) | KTO | 3.989 | perovskite |
| $SrTiO_3$ (100) | STO | 3.905 | perovskite |
| $LaSrGaO_4$ (001) | LSGO | 3.843 | $K_2NiF_4$ |
| $NdGaO_3$ (100) | NGO | 3.838 | perovskite |
| $LaAlO_3$ (100) | LAO | 3.793 | perovskite |
| $LaSrAlO_4$ (001) | LSAO | 3.755 | $K_2NiF_4$ |
| $PrSrAlO_4$ (001) | PSAO | 3.727 | $K_2NiF_4$ |
| $YAlO_3$ (100) | YAO | 3.715 | perovskite |
| $NdSrAlO_4$ (001) | NSAO | 3.712 | $K_2NiF_4$ |
| $NdCaAlO_4$ (001) | NCAO | 3.688 | $K_2NiF_4$ |
| $ZrO_2$(Y) (100) | YSZ | 3.616 | fluorite |
| T'-$La_2CuO_4$ | | 4.005 | $Nd_2CuO_4$ |
| T-$La_2CuO_4$ | | 3.803 | $K_2NiF_4$ |

# 4. PHASE CONTROL OF $La_2CuO_4$

Figure 4 shows the X-ray diffraction (XRD) patterns of LCO films grown at $T_s$ = 550°C on the various substrates. Since the c-axis lattice constant ($c_0$) is distinct between T and T' ($c_0$(T) = 13.15 Å versus $c_0$(T') = 12.55 Å), the phase identification is rather straightforward. The calculated patterns for T and T' are included in Fig. 4(a). Of the films in this figure, the film on KTO is single-phase T', while the films on LSGO and LSAO are single-phase T. On YAO and NCAO, the films are dominantly T' with a trace amount of T. On STO, the film is clearly a mixture of T and T' with some amount of the T*-like (!) phase. On YSZ, judging from the peak positions, the film seems to be single-phase T', although the peak intensity ratios do not agree with the calculated ratios. The $c_0$ values of these films are summarized in Fig. 4(b) together with those of films on other substrates. Because of epitaxial strain,[22] $c_0$ of the T structure is noticeably substrate-dependent: the longest ($c_0$ = 13.25 Å) for LSAO and the shortest ($c_0$ = 13.10 Å) for LSGO. The same investigations were performed at $T_s$ from 450°C to 750°C on all the substrates in Table 2.[23] The

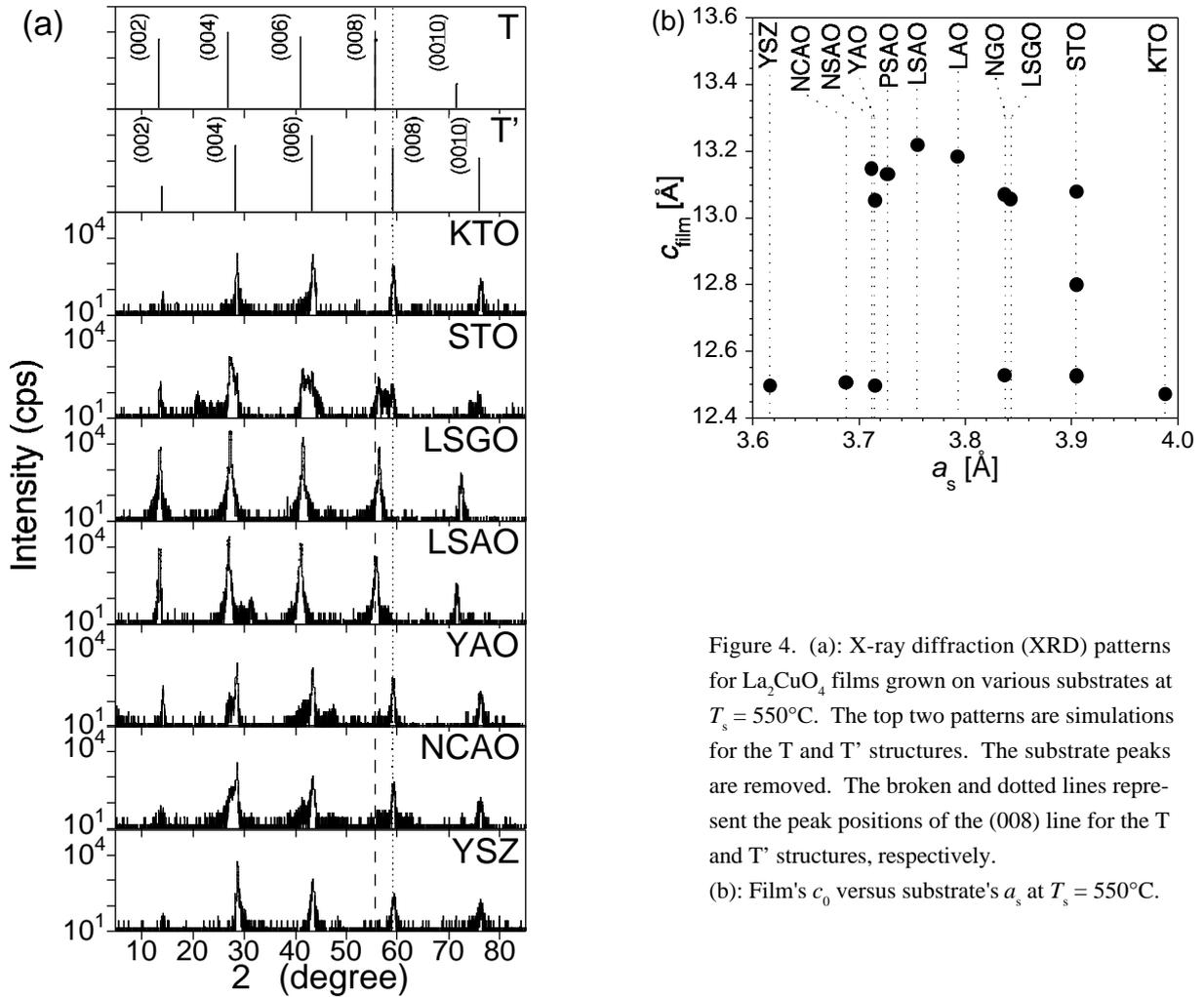

Figure 4. (a): X-ray diffraction (XRD) patterns for $La_2CuO_4$ films grown on various substrates at $T_s$ = 550°C. The top two patterns are simulations for the T and T' structures. The substrate peaks are removed. The broken and dotted lines represent the peak positions of the (008) line for the T and T' structures, respectively.

(b): Film's $c_0$ versus substrate's $a_s$ at $T_s$ = 550°C.

results are summarized in Fig. 5, which shows the phase diagram on the selective stabilization of T versus T' in the $T_s$ - $a_s$ planes.

High $T_s$ (650 ~ 750°C)

The films on most of the substrates are single-phase T. There are three exceptions: KTO, YAO, and YSZ. The films on KTO and YSZ do not show any definite X-ray peak, and the film on YAO is a mixture of T and T'.

Low $T_s$ (475 ~ 625°C)

The films on the T-lattice matched substrates (LSGO, LAO, LSAO, PSAO, and NSAO) are single-phase T. The films on T'-lattice matched KTaO$_3$ and on fluorite YSZ are single-phase T'. The films on other substrates (STO, NGO, YAO, and NCAO) are a mixture of T and T' with T' dominant for lower $T_s$.

From these results, we can see three trends:

*(1) Growth temperature*

High $T_s$ favors T and low $T_s$ favors T'.

*(2) Substrate $a_s$*

Substrates with $a_s$ of 3.70 - 3.85 Å favor T and substrates with $a_s$ of > 3.90 Å or < 3.70 Å favor T' (or disfavor T).

*(3) Substrate crystal structure*

K$_2$NiF$_4$-type substrates favor T and fluorite substrates favor T'. Perovskite substrates are neutral.

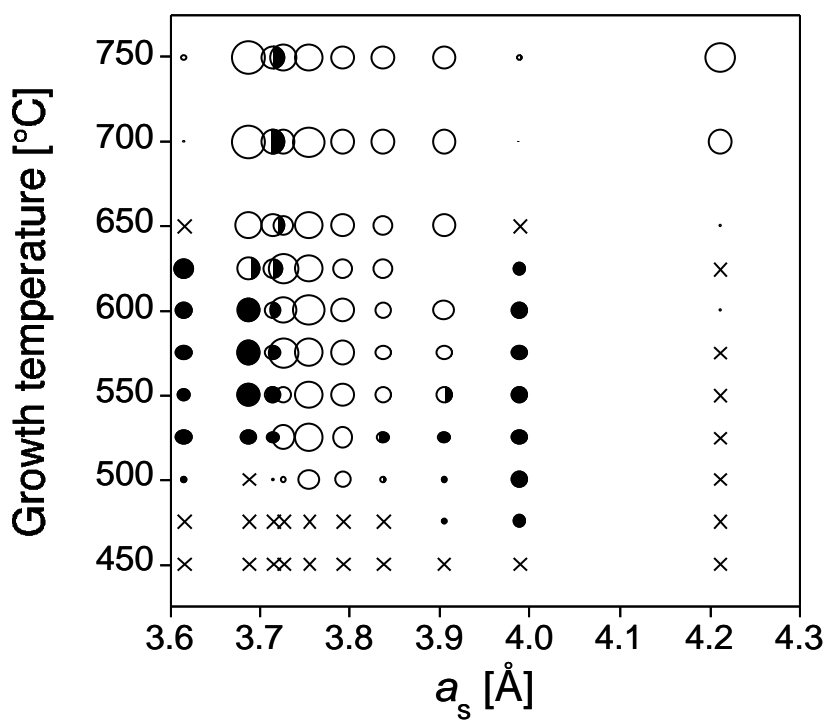

Figure 5. Mapping of the T versus T' phases in the $T_s$ - $a_s$ plane. The crosses represent no phase formation. The open circles represent single-phase T while the filled circles represent single-phase T'. The partially filled circles represent a two-phase mixture. The size (area) of the circles is proportional to the X-ray diffraction peak intensity of the (006) line. For two-phase mixed films, the ratio of the unshaded and the shaded areas represents the ratio of the T and T' peak intensities of the (006) lines.

Next, we briefly compare the physical properties of T-$La_2CuO_4$ and T'-$La_2CuO_4$, which have the same chemical formula but different crystal structures. Figure 6 shows the temperature dependences of resistivity for both phases. The solid lines represent the $\rho$-$T$ curves for the films, which were cooled in vacuum to ambient temperature after film growth. These films do not have excess oxygen but might have slight oxygen deficiencies ($La_2CuO_{4+\delta}$ with $\delta \sim 0$). The broken lines represent those for the films cooled in ozone, which have interstitial excess oxygen ($\delta > 0$). The excess oxygen occupies the tetrahedral site in T, and the apical site in T'. The vacuum-cooled T film has much higher resistivity (several orders of magnitude higher at low temperatures) than the vacuum-cooled T' films. The ozone cooling causes a totally opposite effect on T and T'. The resistivity of the T film reduces by five orders of magnitude at room temperature, from ~ 50 cm to ~ $5 \times 10^{-4}$ cm, indicating that doped holes introduced by excess oxygen are itinerant. Furthermore, the film becomes superconducting. By contrast, the resistivity of the T' film increases, indicating that doped holes are localized. Other physical properties, such as optical spectra, photoemission spectra, also show that T-$La_2CuO_4$ and T'-$La_2CuO_4$ are quite different. The details will be reported elsewhere.[24]

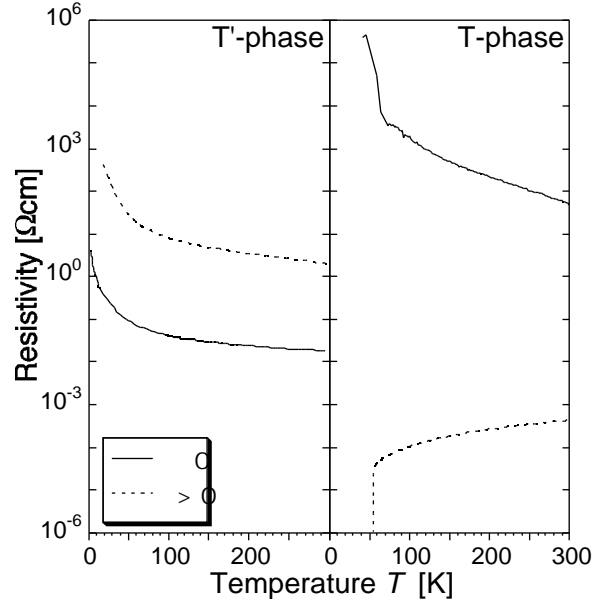

Figure 6. Temperature dependences of resistivity for T- and T'-$La_2CuO_{4+\delta}$ films. The solid lines are for films cooled in vacuum ($\delta \sim 0$) while the broken lines are for films cooled in ozone ($\delta > 0$).

## 5. DOPING

### 5.1 Doping into T-$La_2CuO_4$

Next, we describe our attempts at hole and electron doping into T- and T'-$La_2CuO_4$.[25] First, we present the results on doping into T-$La_2CuO_4$. In the case of hole doping, with LSAO substrates, we attained the best $T_c$ of 45 K for $La_{1.85}Sr_{0.15}CuO_4$, 47 K for $La_{1.85}Ba_{0.15}CuO_4$, and 56 K for $La_2CuO_{4+\delta}$ (LCO+) as shown in Fig. 7. The enhancement of $T_c$ by 20 - 40 % from the bulk values is due to the epitaxial strain effect.

Electron doping into the T structure is empirically impossible by bulk synthesis, as mentioned above. Surprisingly, however, in thin film synthesis, electron doping into T-$La_2CuO_4$ is possible. This is demonstrated in Fig. 8, which shows the doping dependence of $c_0$ for Sr-doped or Ce-doped $La_2CuO_4$ films on LSAO and STO substrates. The $c_0$ value increases with $x$ by hole doping and decreases with $x$ by electron doping. The Ce incorporation into the T lattice can be confirmed by the linear decrease in $c_0$. The solubility limit of Ce into the T structure is $x_c \sim 0.08$ with LSAO substrates and $x_c \sim 0.06$ with STO substrates. At $x_c$, the phase transition to T' takes place as can be seen by the discontinuous change in $c_0$. The slight difference in $c_0$ between films on LSAO and STO, which is especially apparent at hole doping of 0.05 - 0.20, is due to epitaxial strain. We observed that Ce atoms can be incorporated into the T lattice only when the growth temperature is below 680°C. For the growth at temperatures above 700°C, Ce segregates away from the T lattice and $c_0$ stays at the undoped value. The low synthesis temperature seems to be crucial for Ce doping into the T lattice.

Figure 9 shows the $\rho$-$T$ curves of Ce-doped films with single-phase T (at least judging from XRD). Even with doping up to $x \sim 0.06$, the films are semiconducting. The low-temperature resistivity is substantially reduced for $x \sim 0.06$, which may be either due to doping or due to very slight inclusion of T'-$La_2Ce_xCuO_4$, which has much lower resistivity than T-$La_2Ce_xCuO_4$. Anyway, the electron doping up to $x \sim 0.06$ makes T-$La_2CuO_4$ neither metallic nor superconducting. This is in contrast to the situation for hole doping, in which the superconductivity appears at $x \sim 0.05$. Apparently, there exists electron/hole doping asymmetry.

### 5.2 Doping into T'-$La_2CuO_4$

Hole doping into the T' structure is not possible in bulk synthesis. Our preliminary attempt to break this empirical rule by thin film synthesis has not yet been successful. In contrast, electron doping into T'-$La_2CuO_4$ is easily achieved[26, 27] as expected from the tolerance factor consideration in Sec. 2.3. The lattice constants of the resultant films with $x = 0.10$ grown on STO are $a_0 \sim 4.010$ Å and $c_0 \sim 12.45$ Å, which agree with the bulk values.[18] The $a_0$ value is almost doping independent. Substitution of $La^{3+}$ by smaller $Ce^{4+}$ should lead to a decrease in $a_0$, whereas electron doping should lead to an increase of $a_0$. The two effects appear to cancel each other out. On the other hand, $c_0$ decreases almost linearly with $x$ as shown in Fig. 8. The trend lines for this linear relationship for the following three substrates are given as

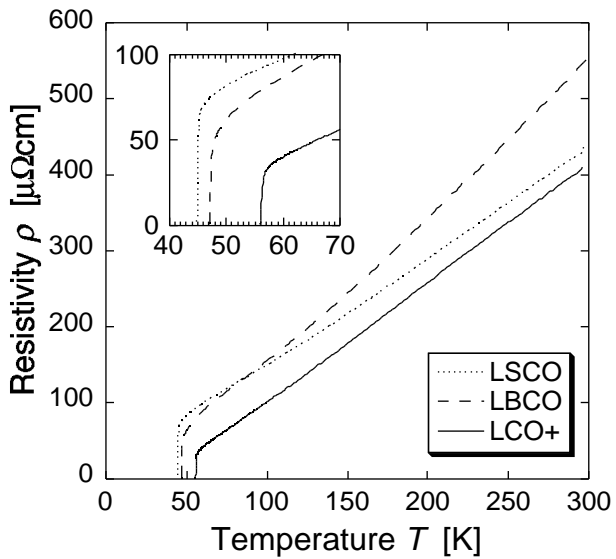

Figure 7. Temperature dependences of resistivity for our best LSCO, LBCO, and LCO+ films. The inset shows superconducting transitions.

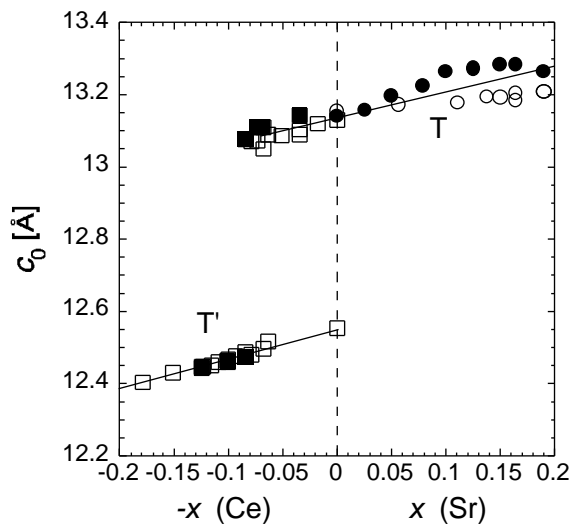

Figure 8. Variation of the $c$-axis lattice constant ($c_0$) for $La_{2-x}M_xCuO_4$ (M = Ce and Sr) as a function of $x$. The open circles and squares are for films on STO substrates while the closed circles and squares are for films on LSAO substrates. In both of the T and T' phases, $c_0$ decreases almost linearly with electron doping. With hole doping in the T phase, $c_0$ increases but shows a strange doping dependence, which is due to epitaxial strain in these films: compressive on LSAO and tensile on STO.

$c_0 = -0.68438 \times x + 12.4980$ for KTO (not shown in Fig. 8),
$c_0 = -0.78950 \times x + 12.5336$ for STO,
$c_0 = -0.77742 \times x + 12.5230$ for LSAO.

The $c_0$ value is the largest with STO, and the shortest with KTO. It indicates that the films on STO have in-plane compressive strain while the films on KTO have in-plane tensile strain. On LSAO, epitaxial strain seems to be relaxed immediately after a few initial unit cells due to too large lattice mismatch.[27]

The typical $\rho$ - $T$ curves are shown in Fig. 10. The optimally doped films have a resistivity value of ~ 300 $\mu$ cm at room temperature and ~ 30 $\mu$ cm at 40 K. The highest $T_c$ is over 30 K at zero resistance. The superconducting transition is very sharp, even sharper than for our best NCCO or PCCO, with the transition width less than 0.5 K. The doping dependence of $T_c$ is summarized in Fig. 11. The transition temperature has a clear maximum at $x$ ~ 0.08 - 0.09. As $x$ is decreased from 0.08, $T_c$ value falls off sharply. At higher doping, $T_c$ falls off more smoothly and superconductivity disappears for $x > $ ~ 0.22. The optimum doping ($x_{opt}$) is ~ 0.08 - 0.09, which is significantly lower than $x_{opt}$ ~ 0.14 for PCCO and $x_{opt}$ ~ 0.15 for NCCO. The reason for this nonuniversality for $x_{opt}$ is not well understood although self-doping due to oxygen deficiencies at the O(2) site might explain the shift of $x_{opt}$ in T'-LCCO.

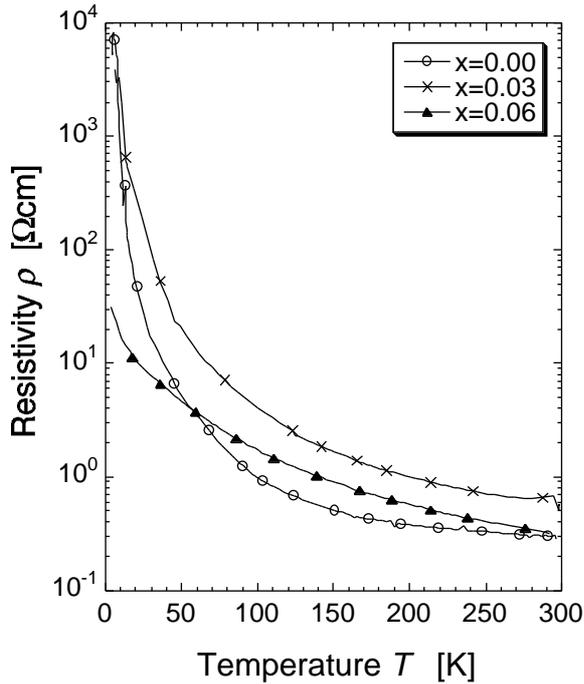

Figure 9. Temperature dependences of resistivity for $La_{2-x}Ce_xCuO_4$ films ($x$ = 0.00 - 0.06) with the T structure grown on LSAO substrates. With Ce doping up to 0.06, the films remain insulating.

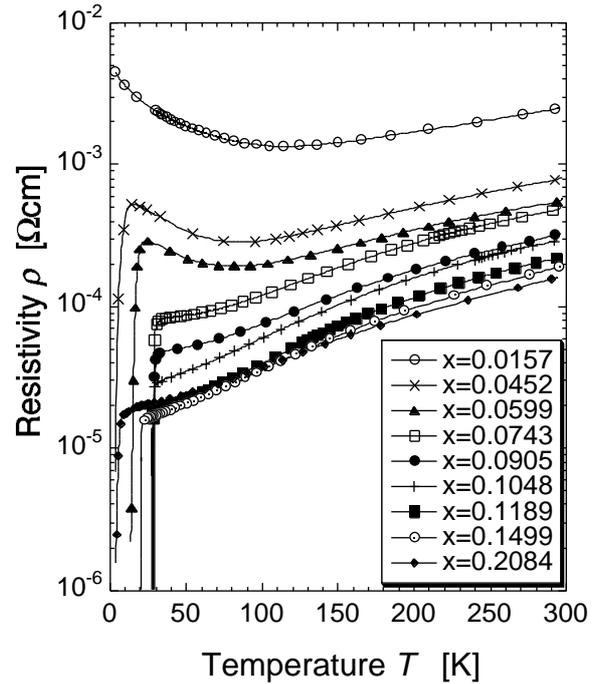

Figure 10. Temperature dependences of resistivity for $La_{2-x}Ce_xCuO_4$ films ($x$ = 0.015 - 0.208) with the T' structure (T'-LCCO) grown on NCAO substrates. The resistivity monotonically decreases with increasing Ce doping.

Figure 12 demonstrates one important trend in the T' family. Figure 12(a) is a plot of the highest superconducting transition temperature versus ionic radius $r_i(Ln^{3+})$. Figure 12(b) is a replot against an in-plane lattice constant ($a_0$). The data for Ln = La, Pr, Nd, and Sm are from our MBE grown films, and those for Ln = Eu and Gd are the bulk values. This obviously demonstrates that a larger $Ln^{3+}$ ion or, equivalently, a larger in-plane lattice constant gives a higher transition temperature. This trend is in accord with the steric effect on $T_c$ versus $a_0$ in the electron-doped cuprates initially pointed out by Markert et al.[28] that $T_c$ is zero for $a_0$ below a critical value ($a_{cr} \sim 3.92$ Å) and increases with $a_0$. This trend can be understood by the following: "Larger $r_i(Ln^{3+})$ or, equivalently, larger $a_0$ facilitates the removal of apical oxygen[29] that is harmful to superconductivity".

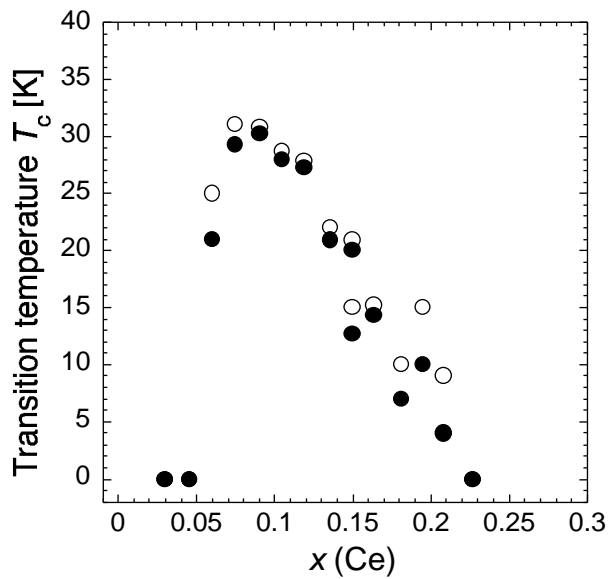

Figure 11. Ce doping dependence of $T_c$ for T'-LCCO films grown on STO substrates. The open and closed circles indicate $T_c$(onset) and $T_c$(end). $T_c$(end) reaches 30 K at $x \sim 0.09$.

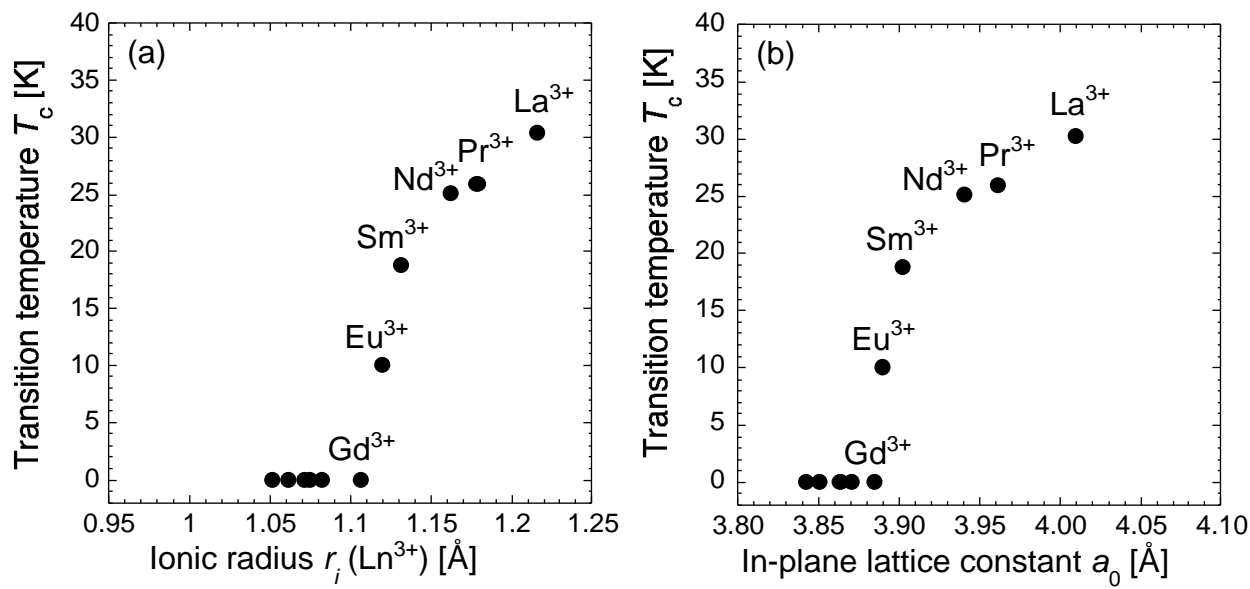

Figure 12. Highest superconducting transition temperature ($T_c$) in the T' family: (a) $T_c$ versus ionic radius ($r_i(Ln^{3+})$) and (b) $T_c$ versus in-plane lattice constant ($a_0$).

# 6. SUMMARY

In summary, we succeeded in the selective stabilization of the T versus T' phase of $La_2CuO_4$ by low-temperature processes based on molecular beam epitaxy. The low growth temperature plays a primary role in stabilizing the T' phase. Furthermore the substrate, depending on $a_s$ and also on the crystal structure, significantly influences the selective stabilization. Our observations can be summarized as follows.

(1) Growth temperature: High $T_s$ favors T and low $T_s$ favors T'.

(2) Substrate $a_s$: Substrates with $a_s$ of 3.70 - 3.85 Å favor T and substrates with $a_s$ of > 3.90 Å or < 3.70 Å favor T' (or disfavor T).

(3) Substrate crystal structure: $K_2NiF_4$-type substrates favor T and fluorite substrates favor T'. Perovskite substrates are neutral.

Electron doping by substituting Ce for La was examined in both T- and T'-$La_2CuO_4$. Single-phase T-$La_{2-x}Ce_xCuO_4$ can be obtained up to $x \sim 0.06$, but neither coherent transport nor superconductivity was achieved. Single-phase T'-$La_{2-x}Ce_xCuO_4$ can be synthesized for a wide range of $x$ ($0 \le x \le 0.35$), and superconductivity with $T_c$(end) ~ 30 K at $x \sim 0.08 - 0.09$ was achieved. In the T' family, there is a clear trend that a larger Ln ion (equivalently larger $a_0$) gives a higher transition temperature. Hole doping into T'-$La_2CuO_4$ has been attempted but not yet with success.

## Acknowledgments

The authors thank Dr. T. Yamada, Dr. H. Yamamoto, and Dr. S. Karimoto for helpful discussions, Dr. S. Yagi for supplying $KTaO_3$ single crystals, Mr. N. Honma for ICP analysis, Dr. H. Takayanagi and Dr. S. Ishihara for their support and encouragement throughout the course of this study.